\documentstyle[12pt]{article}
\begin{document}
\def\theequation{\arabic{section}.\arabic{equation}}
\newcommand{\be}{\begin{equation}}
\newcommand{\ee}{\end{equation}}
\newcommand{\rlnabla}{\nabla^{{}^{{}^{\!\!\!\!\!\!\!\!\!
\longleftrightarrow}}}}
\begin{titlepage}
\title{On the Tail Problem in Cosmology}
\author{Valerio Faraoni \\
\\{\small \it University of Victoria, Department of Physics and
Astronomy,}\\ {\small \it P.O. Box 3055, Victoria, B.C., V8W  3P6
(Canada) }\\
\\ and \\
\\Sebastiano Sonego \\
\\{\small \it Universit\'{e} Libre de Bruxelles, C.P. 231 Campus Plaine
U.L.B.,}\\ {\small \it Boulevard du Triomphe, 1050 Bruxelles (Belgium)} }
\date{}
\maketitle
\thispagestyle{empty}
\vspace*{1truecm}
\begin{abstract}
The tail problem for the propagation of a scalar field is considered in a
cosmological background, taking a Robertson-Walker spacetime as a specific
example. The explicit radial dependence of the general solution of the
Klein-Gordon equation with nonminimal coupling is derived, and the
inapplicability of the standard calculation of the reflection and
transmission coefficients to the study of scattering of waves by the
cosmological curvature is discussed.
\end{abstract}
\vspace*{0.5truecm}
\begin{center}
To appear in {\em Physics Letters A}
\end{center}
\end{titlepage}
\newpage
\section{Introduction}

The Klein-Gordon equation\footnote{We use units in which $ c=1 $, and adopt
the notations of Ref.~\cite{Wald}. The signature of the metric is +2.}
\setcounter{equation}{0}
\be
g^{ab}\nabla_a\nabla_b\Phi -\xi \, R \, \Phi -m^2\,\Phi=0 \;,
                                 \label{1.1}
\ee
where $\xi $ is a numerical constant, $R$ is the curvature scalar, and $m$
is the mass of the scalar field $\Phi$, describes the simplest examples of
relativistic wave propagation in curved spacetime and has consequently
received much attention, especially in relation to the study of quantum
processes \cite{BirrellDavies}. Nevertheless, some interesting features
have not yet been fully investigated: In particular, the physical
consequences of the fact that the solutions of Eq.~(\ref{1.1}), even in the
case in which $\xi=m=0$, generally present {\em tails}
\cite{Hadamard}--\cite{SV} (i.e., propagation in the interior of the light
cone), has been considered only in the context of radiation emitted by
compact objects \cite{Price}--\cite{NovikovFrolov}.

The formation of tails when $\xi=m=0$ can be understood as an effect of the
backscattering of curvature on the waves; this is schematically represented
in Fig.~1, which represents a portion of spacetime in null coordinates
$ u $, $ v $. The curvature is nonzero only in the shaded region, and
scatters waves analogously to what happens under the action of a potential.
The ray 1, which does not enter the curved region, remains unperturbed,
whereas the ray 2 can be decomposed into a transmitted (3) and a reflected
(4) part. Nevertheless we emphasize that such a representation is purely
pictorial and does not properly describe many features of the real
phenomenon, which has a continuous, rather than localized character. A
better, although still schematic, picture is given in Fig.~2. As in the
usual scattering processes reflection is due to the inhomogeneities of the
potential, so here it is caused by the curvature. In fact, a region of
spacetime containing a homogeneous gravitational field is flat, and it is
known that waves are not diffused by a flat four dimensional background
\cite{Hadamard,Kunzle,Friedlander}.

As said above, this effect has been investigated for waves emitted by a
compact source, by calculating explicitly the reflection and transmission
coefficients, and has been found to be significant only in the induction
zone and for long wavelengths, becoming negligible at large distances from
the source and at late times, so that it is irrelevant for the problem of
wave propagation in presence of astrophysical objects \cite{Thorne}. These
general results can be qualitatively understood by remembering that scattering
is non-negligible only in the regions where curvature is appreciable. Since
the curvature of spacetimes associated to localized matter distributions
drops off quickly as spatial infinity is approached, the region of
scattering is localized near the source, as in Fig.~3. Moreover, since
the ``potential barrier'' turns out not to be very high, most of the
radiation escapes in the region where the curvature is negligible, and
where it is no more backscattered. Reflection is appreciable only for long
wavelengths.

These considerations do not apply to spacetimes relevant as cosmological
models, in which curvature is present over large scales and scattering
could, in principle, take place everywhere (see Fig.~4). It is thus natural
to ask whether in these cases the process can eventually ``convey'' a
relevant part of radiation within the light cone; this constitutes the
so-called cosmological {\em tail problem} \cite{EllisSciama}. Whereas
{\em qualitative} criteria for the existence of tails have been already
formulated and applied to special cases \cite{Hadamard}--\cite{SV}, no
general {\em quantitative} characterization of the phenomenon, that could
solve the problem above, seems to have been suggested up to now. A proposal
of this kind, attempting to quantify the physical relevance of wave tails,
will be described in a forthcoming article \cite{noi}. In this paper we
restrict ourselves to present (Sec.~2) a solution of Eq.~(\ref{1.1}) in a
form which can be useful for later references, and to point out (Sec.~3)
the inapplicability of methods based on the calculation of reflection and
transmission coefficients to investigate the occurrence of tails in a
cosmological background. Sec.~4 contains some general remarks about the
implications of this result.

All the calculations will be performed in a
Friedmann-Lemaitre-Robertson-Walker (FLRW) universe, with
metric\footnote{$\eta$ is the so-called {\em conformal time}
\cite{BirrellDavies}.}
\be
ds^{2}=a^{2}(\eta)\left[-d\eta^{2}+ d\chi^{2}+f^{2}( \chi ) \left(
d\theta^{2}+ \sin^{2} \theta \, d\varphi^{2} \right) \right]\;,
                                 \label{1.2}
\ee
where
\be
f( \chi )=\left\{ \begin{array}{cllll}
\sinh \chi & \,\,\,\,\,\,\chi\in(0,+\infty)& \;\;\; \mbox{if}
\,\,\,\,\,\, & K=-1 & \nonumber \\
\chi & \,\,\,\,\,\,\chi\in(0,+\infty)&\;\;\; \mbox{if}
\,\,\,\, & K=0 & \nonumber \\
\sin \chi & \,\,\,\,\,\,\chi\in(0,\pi)&\;\;\; \mbox{if} \,\,\,\,\,\,
& K=+1 \; , & \nonumber  \end{array} \right.
                                 \label{1.3}
\ee
and $a(\eta)$ is the scale factor. Although the general solution of
Eq.~(\ref{1.1}) in a FLRW spacetime is well-known in an implicit form
\cite{ParkerFulling,BirrellDavies}, the explicit dependence on $\chi$ has
been worked out, to the authors' knowledge, only in the case $K=+1$
\cite{Ford} (the case $K=0$ is trivial). As far as the study of tails is
concerned, however, this is the most intricate case, because the
phenomenon, if present, is complicated by the possibility that radiation
travels more
than once through closed spatial sections. For this reason, and since it
represents the only one not yet considered in the literature, we shall work
out explicitly only the case $K=-1$.

\section{General solution}

Equation (\ref{1.1}) can be rewritten in the form\setcounter{equation}{0}
\be
\frac{1}{\sqrt{-g}}\frac{\partial}{\partial x^a}\left( \sqrt{-g} \,
g^{ab}\,
\frac{\partial\Phi}{\partial x^b} \right) -\xi\,R\,\Phi-m^2\,\Phi=0 \;.
                              \label{eqforphi}
\ee
In the FLRW background defined by Eqs.~(\ref{1.2}) and (\ref{1.3}),
Eq.~(\ref{eqforphi}) becomes
\begin{eqnarray}
\lefteqn{\frac{1}{f^{2}}\left[ \frac{\partial}{\partial\chi}\left(f^{2}
\frac{\partial\Phi}{\partial\chi} \right)+\frac{1}{\sin \theta }\,
\frac{\partial}{\partial\theta}\left( \sin \theta \, \frac{\partial\Phi}
{\partial\theta} \right)+ \frac{1}{\sin^{2}\theta}\,
\frac{\partial^2\Phi}{\partial\varphi^2} \right]} \nonumber \\
&&-\,\frac{1}{a^2}\,\frac{\partial}{\partial \eta}\left(a^{2}\,
\frac{\partial\Phi} {\partial \eta}\right)-a^2\,\xi\,R\,\Phi-a^2\,m^2\,
\Phi=0\;,
                             \label{waveeq}
\end{eqnarray}
where now
\be
R=6\left( \frac{\ddot{a}}{a^3}+\frac{K}{a^2}\right)\;,
                                 \label{curv}
\ee
a dot denoting derivative with respect to $\eta$. Separation of time and
space variables,
\be
\Phi (\eta,\chi ,\theta ,\varphi)=T(\eta)S( \chi ,\theta ,\varphi ) \; ,
                                  \ee
leads to
\be
\frac{1}{a^4}\,\frac{d}{d\eta}\left(a^{2}\,\frac{dT}{d\eta}\right)+
\xi\,R\,T+m^2\,T+\frac{k}{a^2}\, T=0
                                \label{2.5}
\ee
and
\be
\frac{\partial}{\partial\chi}\left(f^{2}\frac{\partial S}{\partial\chi}
\right)+\frac{1}{\sin\theta}\,\frac{\partial}{\partial\theta}\left( \sin
\theta \,\frac{\partial S}{\partial\theta} \right)+ \frac{1}{\sin^{2}
\theta
}\,\frac{\partial^2 S}{\partial\varphi^2} +kSf^{2} =0\;,
                                \label{2.6}
\ee
where $k$ is a separation constant. Equation~(\ref{2.5}) is in general
difficult to solve, but it becomes very simple when considering a static
universe \cite{Ford,BirrellDavies}, in which $a=$const.~\footnote{The
general
solution of Eq.~(\ref{2.5}) is given in Ref.~\cite{FordParker} also in the
case $ a(\eta) \propto \eta^\alpha $ ($\alpha\neq 1/2 $).} In this case the
solutions are the usual exponentials, $\exp(\pm i\omega \eta)$, where
\be
\omega^2=k+a^2\,m^2+6 \,\xi\,K \;.
                                  \label{omega}
\ee

Further separation of radial and angular coordinates in Eq.~(\ref{2.6}),
\be
S( \chi ,\theta ,\varphi )=X( \chi ) \, Y( \theta ,\varphi )\;,
\ee
leads to the usual equation for spherical harmonics \cite{Lebedev,Landau}
$Y_{lm}(\theta ,\varphi )$ and to
\be
\frac{d}{d\chi}\left(f^{2}\,\frac{dX_{l}}{d\chi}\right)+kf^{2}X_{l}=
l(l+1)X_{l}\;.
                                 \label{2.8}
\ee
Setting $ \Psi_l( \chi ) \equiv f( \chi ) X_{l}( \chi ) $ and using
Eq.~(\ref{1.3}), Eq.~(\ref{2.8}) takes the form of a one-dimensional
Schr\"{o}dinger equation:
\be
\frac{d^2 \Psi_l}{d \chi^2}+\left[ 2E-\frac{l(l+1)}{f^2( \chi)} \right]
\Psi_l=0 \; ,
                                    \label{27}
\ee
where
\be
2E \equiv k+K \; .
                                    \label{28}
\ee

The explicit solution of Eq.~(\ref{27}) in the case $K=+1$ has been already
given by Ford \cite{Ford} in terms of Gegenbauer functions. For $K=0$, on
the other hand, Eq.~(\ref{27}) becomes
\be
\frac{d^2 \Psi_l}{d \chi^2}+\left[ 2E-\frac{l(l+1)}{\chi^2} \right]
\Psi_l=0 \; ,
                                  \label{210}
\ee
which is familiar from the quantum mechanical description of a free
particle
in spherical coordinates (see, e.g., Ref.~\cite{Landau}). Its general
solution is a linear combination of
\be
\Psi_l^{(1,2)}(\chi)\equiv
\sqrt{\chi}\,H_{l+1/2}^{(1,2)}(\sqrt{2E}\,\chi)\;,
                                  \label{boh}
\ee
where $E>0$ and $H_{l+1/2}^{(1,2)}$ are the Hankel functions
\cite{Lebedev,Landau}.

In the case $ K=-1, $ Eq.~(\ref{27}) allows the eigenvalue $ E $ to assume
any positive value. It is convenient to define
\be
 2E \equiv p^2 \; ,
                                   \label{212}
\ee
with $ p>0 $; then
\be
\frac{d^2 \Psi_l}{d \chi^2}+\left[ p^2-\frac{l(l+1)}{\sinh^2 \chi} \right]
\Psi_l=0 \; .
                                   \label{31}
\ee
Writing
\be
\Psi_l( \chi) \equiv e^{ip\chi} \, \phi_{l}( \chi) \; ,
                                   \label{32}
\ee
Eq.~(\ref{31}) transforms in the following equation for $ \phi_l $:
\be
\frac{d^2 \phi_l}{d \chi^2}+2\, ip\, \frac{d\phi_l}{d\chi}-
\frac{l(l+1)}{\sinh^2 \chi} \, \phi_l=0 \; .
                                   \label{33}
\ee
It is now useful to define the new variable $ z \in \left( -\infty, 0
\right) $ as
\be
 z\equiv \frac{1}{2} \, (1-\coth \chi )=\frac{1}{1-e^{2\chi}} \;.
                                   \label{34}
\ee
In terms of $ z $, Eq.~(\ref{33}) takes the form
\be
z\, (1-z) \, \frac{d^2 \phi_l}{dz^2}+\left(1-ip-2z \right)
\frac{d\phi_l}{dz}+l(l+1) \phi_l=0 \; ,
                                  \label{35}
\ee
which is immediately recognized as an hypergeometric equation
\cite{Lebedev,Landau} with coefficients
\begin{eqnarray}
\alpha & = & -l \nonumber \\
\beta & = & l+1                   \label{36} \\
\gamma & = & 1-ip \; . \nonumber
\end{eqnarray}
This equation admits two independent solutions \cite{Lebedev,Landau},
\be
\phi_l^{(1)}=F( \alpha,\beta,\gamma;z)=F(-l,l+1,1-ip;z) \;,
                                  \label{37}
\ee
and
\begin{eqnarray}
\phi_l^{(2)}=z^{1-\gamma}(1-z)^{\gamma-\alpha-\beta}F(1-\beta,\, 1-\alpha,
\,2-\gamma;z)= \nonumber \\
=(-1)^{ip}e^{-2ip\chi} F(-l,l+1,1+ip;z) \; ,
                                  \label{38}
\end{eqnarray}
where $ F $ denotes the hypergeometric function. Correspondingly,
Eq.~(\ref{32}) gives the two independent solutions of Eq.~(\ref{31}) (an
irrelevant factor $(-1)^{ip}$ has been dropped in getting Eq.~(\ref{310})
from Eq.~(\ref{38})):
\begin{eqnarray}
\Psi_l^{(1)}( \chi)=e^{ip\chi}F(-l,l+1,1-ip;z( \chi)) \; ;
                                    \label{39}   \\
\Psi_l^{(2)}( \chi)=e^{-ip\chi}F(-l,l+1,1+ip;z( \chi)) \; .
                                   \label{310}
\end{eqnarray}
The general solution of Eq.~(\ref{31}) is thus
\be
\Psi_l ( \chi)=A_l(p) \,\Psi_l^{(1)}( \chi)+B_l(p)\,\Psi_l^{(2)}( \chi) \;
,
                                  \label{311}
\ee
with $ A_l$ and $B_l$ arbitrary complex functions of $p$.

Notice that, since $ \alpha $ is integer, the series defining $ F $
terminates, and therefore $ F $ reduces to a polynomial of degree $ l $ in
$ z $:
\be
F(-l,l+1,1\mp ip;z)=\sum_{n=0}^{l} \left( \begin{array}{c} l+n \nonumber \\

l \nonumber \end{array} \right) \frac {(-l)_n}{(1\mp ip)_n} \, z^n \; ,
                                 \label{315}
\ee
where $(\lambda)_n$ is defined \cite{Lebedev}, for an arbitrary complex
number $ \lambda $, by
\be
( \lambda)_n \equiv \lambda( \lambda+1)\cdots (\lambda+n-1)=
\frac{\Gamma(\lambda+n)}{\Gamma( \lambda)} \; ,
                                 \label{316}
\ee
and $ ( \lambda)_0 \equiv 1 $. This result is useful if one wants to
compare the exact solution~(\ref{311}) with the asymptotic one for $
\chi\rightarrow 0 $.

\section{Asymptotic character of the solution}

In this section we argue that the concept of reflection and transmission
coefficients is inapplicable to the study of tails in cosmological
backgrounds. To this purpose, we shall restrict the discussion to the case
$K=-1$ and $a=$const. For $K=+1$, due to the existence of closed spatial
sections of spacetime, the transmitted radiation might be allowed to pass
more than once through a given point of space, superposing to the fraction
of radiation which is possibly reflected. For non-constant $a(\eta)$, the
simple exponentials $\exp (\pm ip\chi)$ might not correspond to outgoing
and ingoing waves. In both cases the treatment would suffer from unnecessary
complications.

The presence of tails is characterized by the fact that a pulse of
radiation
emitted at $\chi=0$ is partially backscattered. A stationary emission must
therefore correspond to a solution of Eq.~(\ref{waveeq}) which for $\chi
\rightarrow 0$ contains incoming (reflected) as well as outgoing (emitted)
radiation, whereas for $\chi \rightarrow +\infty$ only outgoing
(transmitted) waves are present. Therefore, we must impose to the general
solution~(\ref{311}) a boundary condition that corresponds to the absence
of incoming waves at $\chi \rightarrow +\infty$. This is easily accomplished
by
noticing that for $\chi \rightarrow +\infty$ one has $z \rightarrow 0$ and
$
F\rightarrow 1$. Hence, the asymptotic form of $\Psi_l^{(1)}$ and
$\Psi_l^{(2)}$ is\setcounter{equation}{0}
\begin{eqnarray}
\Psi_l^{(1)}( \chi \rightarrow +\infty ) \approx e^{ip\chi} \; ,
                                   \label{312}  \\
\Psi_l^{(2)}( \chi \rightarrow +\infty ) \approx e^{-ip\chi} \; ,
                                   \label{313}
\end{eqnarray}
which correspond, respectively, to outgoing and ingoing waves. The required
boundary condition is therefore $ B_{l}=0 $, leading to
\be
\Psi_l( \chi)=A_l(p) \, e^{ip\chi} F(-l,l+1,1-ip;z( \chi))
                                  \label{314}
\ee
as the specific solution of the problem.

A calculation of the reflection and transmission coefficients along the
same lines of that performed in Refs.~\cite{Price}--\cite{NovikovFrolov}
would now require to expand $\Psi_l(\chi)$ as
\be
\Psi_l( \chi)=A_l^{(+)} \, \Psi_l^{(+)} ( \chi)+ A_l^{(-)} \, \Psi_l^{(-)}
( \chi) \;,
                                  \label{G6}
\ee
where $ \Psi_l^{( \pm)} $ are independent solutions of Eq.~(\ref{31})
corresponding to waves which are purely outgoing $(+) $ and ingoing $ (-) $
for $ \chi \rightarrow 0 $. Unfortunately, such solutions do not exist for
$ l \neq 0 $ if we require purely ingoing/outgoing waves to have the
form\footnote{The following results are conditioned by the use of this
characterization, which appears rather restrictive. However, other
definitions do not seem sharp enough in selecting waves which can be
considered as {\em purely} ingoing or outgoing. The formulation of
a more general characterization would nevertheless constitute an
important progress as far as the calculation of reflection and
transmission coefficients is concerned.} $\exp[-i(\omega\eta\pm p\chi)]$ as
$\chi\rightarrow 0$. In fact, these simple exponentials are not solutions
of the wave equation near the origin. This can be qualitatively understood by
observing that in the  neighbourhood of $ \chi = 0 $ the centrifugal
potential $ l(l+1) /\chi^2 $  varies extremely rapidly, preventing the
modes
$\exp(\pm ip\chi)$ from occurring  separately. As a consequence, one cannot
select solutions of Eq.~(\ref{31})  which correspond to waves that near the
origin have a purely ingoing or outgoing  behaviour\footnote{It is easy to
check that the same conclusion holds for a  generic solution of the wave
equation that contains components with  $l\neq 0$. An analogy with the
behaviour of particles can be drawn starting  from the similarity of
Eq.~(\ref{31}) with a Schr\"{o}dinger equation in a  central potential. The
intuitive explanation is that particles with nonzero  angular momentum
cannot ``enter in'' or ``exit from'' the centre  $ \chi=0 $.}.

Formally, this can be realized by characterizing these solutions as
asymptotic eigenfunctions of the operator $ \hat{P}_{\chi}
\equiv -i d/d\chi $, i.e.,
\be
-i \,\, \frac{d \Psi_l^{( \pm)}}{d\chi}=\pm \alpha \, \Psi_l^{( \pm)}+
\left( \mbox{higher powers of $ \chi$} \right) \;,
                                   \label{G8}
\ee
with $ \alpha > 0 $. Since for $l\neq 0$
\be
\left[ \hat{H}_l, \hat{P}_{\chi} \right] =O( \chi^{-3}) \neq \hat{0} \;,
                                    \label{G9}
\ee
where
\be
\hat{H}_l \equiv -\, \frac{1}{2} \, \frac{d^2 }{d\chi^2} +\frac{l(l+1)}{2
\,
\sinh^2 \chi } \;,
                                  \label{G10}
\ee
it follows that there are no common eigenstates of the operators
$ \hat{H}_l $ and $ \hat{P}_{\chi} $, i.e., there are no solutions of
Eq.~(\ref{31}) which satisfy the asymptotic condition (\ref{G8}).

This conclusion can be also checked explicitly as follows. Looking for
solutions of Eq.~(\ref{31}) of the form
\be
\Psi_l(\chi)=\chi^m+O(\chi^{m+1}) \;,
                                  \label{G2}
\ee
one finds immediately that either $ m=-l $ or $ m=l+1 $. The general
solution of Eq.~(\ref{31}) can therefore be written as
\be
\Psi_l( \chi)=C_l \, f_l( \chi)+D_l \, g_l( \chi) \;,
                                  \label{G3}
\ee
with $ C_l$ and $D_l$ arbitrary complex numbers and $ f_l, \, g_l $ two
particular solutions which behave, for $ \chi \rightarrow 0$, as
\be
f_l( \chi)=\frac{1}{\chi^l}+O( \chi^{1-l}) \;,
                                  \label{G4}
\ee
and
\be
g_l( \chi)=\chi^{l+1}+O( \chi^{l+2}) \;.
                                  \label{G5}
\ee
The coefficients $ C_l $ and $ D_l $ for any particular solution $ \Psi_l $
can be obtained by comparing Eqs.~(\ref{G3})--(\ref{G5}) with the expansion
of $ \Psi_l ( \chi) $ for $ \chi \rightarrow 0 $. In particular, for
$ \Psi_l^{( \pm)} $ one can write
\be
\Psi_l^{( \pm)}( \chi)=C_l^{( \pm)} \, f_l( \chi)+D_l^{( \pm)}
\, g_l( \chi) \;,
                                  \label{G7}
\ee
where the coefficients $ C_l^{( \pm)} $ and $ D_l^{( \pm)} $ should be
chosen in such a way as to guarantee the prescribed character (purely
outgoing or ingoing) of the solutions $ \Psi_l^{( \pm)} $ for $ \chi
\rightarrow 0 $. Substituting Eq.~(\ref{G7}) into (\ref{G8}) and requiring
the coefficients of the leading powers of $ \chi $ to vanish, one obtains
the trivial result $ C_l^{( \pm)}=D_l^{( \pm)}=0 $.

The non-existence of solutions of Eq.~(\ref{31}) corresponding to
waves which are purely ingoing
or outgoing in the region $ \chi \rightarrow 0 $, prohibits one to define
reflection and transmission coefficients as in the usual treatments of
scattering problems. It seems that no method has been developed in the
literature to deal with similar situations. The case considered here is
quite different from the simpler ones arising when studying diffusion of
waves by Schwarzschild black holes, where a coordinate transformation can
be found for which the corresponding one dimensional Schr\"{o}dinger
problem involves a finite, localized potential barrier, and nontrivial
asymptotic eigenfunctions of momentum can be found
\cite{Price}--\cite{NovikovFrolov}.

\section{General remarks}

The most appropriate treatment of the tail problem would be the explicit
determination of the reflection coefficient describing backscattering by
the cosmological curvature, and characterizing quantitatively the fraction of
radiation which does not propagate along the light cone. The impossibility
of carrying on this approach leads one to look for alternative ways of
studying the phenomenon. A straightforward idea could be to pursue the
formal analogy between the tail problem and quantum scattering, by
observing that Eq~(\ref{31}) can be thought of as the radial part of a
stationary Schr\"{o}dinger equation describing a particle with Hamiltonian
operator\setcounter{equation}{0}
\be
\hat{H}=\frac{1}{2}\,\hat{\bf p}^2+\frac{1}{2}\, V(\chi) \,\hat{\bf L}^2\;,
                                \label{4.1}
\ee
where $\hat{\bf p}$ and $\hat{\bf L}$ are, respectively, the linear and
angular momentum operators in a fictitious three-dimensional Euclidean
space in which $\chi$ plays the role of an ordinary radial coordinate,
$\Psi_l(\chi)/\chi$ is the radial part of the $l$-th component of the
Schr\"{o}dinger wave function, and
\be
V(\chi)\equiv\frac{1}{\sinh^2\chi}-\frac{1}{\chi^2}\;.
                                   \label{4.2}
\ee
One possibility would then be to compute the cross section for this quantum
mechanical system. Although such a calculation could be carried on, it
cannot however be regarded as a satisfactory solution to our specific
physical problem, which concerns waves emitted from $ \chi=0 $ rather than
incoming from infinity. A satisfactory study of the subject seems thus to
require a radically different approach. In Ref.~\cite{noi}, a technique
will be presented which characterizes tails in terms of the ratio between
their energy content and the total energy of the field.

The impossibility of defining reflection and transmission coefficients is
not the only reason (although obviously a compelling one) to look for a
different characterization of tails. Even if such coefficients could be
defined, in fact, they would not account in a reliable way for the fraction
of radiation propagating inside the light cone. This can be realized by
noticing that a crucial part of the information about the character of wave
propagation is exclusively contained in Eq.~(\ref{2.5}), whereas the
calculation of the reflection and transmission coefficients involves only
Eq.~(\ref{2.8}). For example, as far as Eq.~(\ref{2.8}) (and, hence, the
reflection and transmission coefficients) is concerned, all the values of
$\xi$ and $m$ are equivalent, which is certainly not the case since, e.g.,
the choice $m=0$, $\xi=1/6$ leads to no tails in every conformally flat
spacetime \cite{Friedlander}. A way to take into account the time
dependence as well is to perform the separation
\be
\label{4.3}
\Phi ( \eta, \chi,\theta, \varphi )=
\frac{\psi ( \eta , \chi )}{f( \chi )} \,\, Y( \theta, \varphi )
\ee
in Eq.~(\ref{waveeq}), getting
\be
\label{4.4}
\frac{\partial^2 \psi_l}{\partial \chi^2}-\frac{\partial^2 \psi_l}{\partial
\eta^2}-\left[ 1-6\, \xi +a^2 m^2 +\frac{l(l+1)}{\sinh^2 \chi} \right]
\psi_l =0 \; .
\ee
Under the coordinate transformations \cite{CT86}
\be
\left\{ \begin{array}{cllll}
u = \frac{1}{2}\, \left( \eta -\chi \right) & & & \nonumber \\
v= \frac{1}{2} \left( \eta+\chi \right) \; , & & & \nonumber \\
\end{array} \right.
\label{4.5}
\ee
\be
\left\{ \begin{array}{cllll}
\bar{u} = \tanh u & & & \nonumber \\
\bar{v}= \tanh v \; , & & & \nonumber \\
\end{array} \right.
\label{4.6}
\ee
\be
\left\{ \begin{array}{cllll}
\bar{\eta} = \bar{v}+\bar{u} & & & \nonumber \\
\bar{\chi} = \bar{v}-\bar{u} \; , & & & \nonumber \\
\end{array} \right.
\label{4.7}
\ee
Eq.~(\ref{4.4}) becomes
\be
\label{4.8}
\frac{\partial^2 \psi_l}{\partial \bar{\chi}^2}-\frac{\partial^2
\psi_l}{\partial
\bar{\eta}^2}-\left[ \frac{1-6\,\xi +a^2 m^2}{(1-\bar{u}^2)(
1-\bar{v}^2)} +\frac{l(l+1)}{\bar{\chi}^2} \right] \psi_l =0 \; .
\ee
If
\be
\label{4.9}
1-6\, \xi +a^2m^2=0 \; ,
\ee
Eq.~(\ref{4.8}), and consequently Eq.~(\ref{eqforphi}), is clearly
tail-free, since it corresponds formally to wave propagation in flat
spacetime\footnote{We thank an anonimous referee for pointing out this
consideration.}. This agrees with previous literature
\cite{KundtNewman,CT86}, and can be related to the known fact that $
\sinh^{-2} \chi $ is a so-called reflectionless potential
\cite{Bargmann,KayMoses}. However, both this argument and previous
works are not relevant when Eq.~(\ref{4.9}) does not hold -- for
example if $ m=\xi =0 $.

In spite of the deficiencies of the treatment, one feature of the
phenomenon emerges nevertheless clearly: The tail problem for radiation in
FLRW spacetimes regards in general only waves whose wavelengths are at
least of order $ a(\eta) $. This can be seen explicitly for the case
of a static universe with $ K=-1 $, in which Eqs.~(\ref{omega}), (\ref{28})
and (\ref{212}) give, for the case $m=0 $,
\be
\omega^{2}=p^2+1-6\,\xi \; ,
                                 \label{61}
\ee
that can be regarded as the dispersion relation for the waves. From
Eq.~(\ref{61}) it follows that, except in the very special case $\xi=1/6 $,
only components with frequency and wave number\footnote{The frequency and
wave number measured by a fundamental observer differ from $\omega$ and $p$
by a factor $a^{-1}$.} smaller than (or comparable to) $ a^{-1} $ are
subject to diffusion. Since tails would then regard only a very small,
extreme part of the spectrum of the radiation content of the universe, and
since the variability of these waves on scales of order $ a $ is too slow
to be detected, it appears that the effect is hardly observable. However,
no conclusion can be drawn without a careful investigation; in fact
radiation of very long wavelengths can possibly lead to physically relevant
effects (see, e.g., Ref.~\cite{HochbergKephart}).

Scalar fields satisfying a wave equation are considered in the
inflationary models of the early universe (see, e.g.,
\cite{KolbTurner} and references therein). The tail problem for the
inflaton regards length scales relevant
for cosmology, so it could possibly have some importance in problems
connected to the physics of the early universe. This is however beyond
the scope of the present work.

\section*{Acknowledgments}

It is a pleasure to thank Dr. M. Bruni and Professor G.~F.~R. Ellis for
helpful discussions, and an anonimous referee for stimulating several
improvements in the presentation. This work was supported by the Italian
Ministero dell'Universit\`a e della Ricerca Scientifica e Tecnologica and
by the Directorate--General for Science, Research and Development of the
Commission of the European Communities under contract n.~CI1-0540-M(TT).

{\small 
\end{document}